\documentstyle[prd,aps,epsf,epsfig,twocolumn]{revtex} %c2
\begin{document}
\onecolumn

\draft

\twocolumn[\hsize\textwidth\columnwidth\hsize\csname@twocolumnfalse\endcsname%c2

%:::::::::::::::::::::::::::::::::::::::::::::::::::::::::::::::::::::::
%
\title{
Evolution of the social network of scientific collaborations}
%:::::::::::::::::::::::::::::::::::::::::::::::::::::::::::::::::::::::
\author
{A.L. Barab\'asi$^{1,2}$, H. Jeong$^{1}$, Z. N\'eda$^{1,2,}$\cite{NZ}, E. Ravasz$^{1}$,
 A. Schubert$^3$, T. Vicsek$^{2,4}$ }
\address{
$^1$Department of Physics, University of Notre Dame, Notre Dame,
IN 46556, USA \\
$^2$ Collegium Budapest, Institute of Advanced Study, Budapest, Hungary \\
$^3$ Bibliometric Service, Library of the Hungarian
Academy of Sciences, Budapest, Hungary     \\
$^4$Department of Biological Physics, E\"otv\"os Lor\'and University,
Budapest, Hungary
}

\maketitle
\centerline{\small (Last revised \today)}

\begin{abstract}
	The co-authorship network of
scientists represents a prototype of complex evolving networks.
	In addition, it offers  one of the most
extensive database to date on social networks.
	By mapping the electronic database
containing all relevant journals in   mathematics and neuro-science
for an  eight-year period (1991-98), we infer the dynamic and the
structural mechanisms that govern the evolution and topology of this complex system.
	Three complementary approaches allow us to obtain a detailed characterization.
	First, empirical measurements allow us to uncover the topological measures that characterize the network
at a given moment, as well as the time evolution of these quantities.
	The results indicate that the network is scale-free, and that the network evolution
 is governed by preferential attachment, affecting both internal and external links.
	However, in contrast with most model predictions the average degree increases
in time, and the node separation decreases.
	 Second, we propose a simple model that captures the network's
time evolution.
	In some limits the model can be solved analytically, predicting a
two-regime scaling in agreement with the measurements.
	Third, numerical simulations are used to uncover the behavior of quantities that
 could not be predicted analytically.
	The combined numerical and analytical results underline the important role internal
links play in determining the observed scaling behavior and network topology.
	  The results and  methodologies
developed in the context of the co-authorship network could
be useful for a systematic study of other complex
evolving networks as well, such as the world wide web, Internet, or other social networks.

\end{abstract}

\pacs{PACS numbers: 89.65.-s, 89.75.-k, 05.10.-a}
\vspace{2pc}
]%c2

\vspace{1cm}

\narrowtext

%:::::::::::::::::::::::::::::::::::::::::::::::::::::::::::::::::::::::
% Text begins:
%:::::::::::::::::::::::::::::::::::::::::::::::::::::::::::::::::::::::

\section{Introduction}

One of the most prolific mathematicians of all time, Paul Erd\H{o}s has written
over 1400 papers with over 500 co-authors. This unparalleled productivity
inspired the concept of the
Erd\H{o}s number, which is defined to be one for his many
co-authors, two for the co-authors
of his co-authors and so on. The tightly interconnected nature of the
scientific community is reflected by the conjecture that all publishing
mathematicians, as well as many physicists and economists have rather small
Erd\H{o}s numbers \cite{erdosnr}.  Besides the immediate
interest for scientometrics, the co-authorship networks  is
of general interest for understanding the
topological and dynamical laws governing complex networks
\cite{watts1,alb1,watts2,dorog1,d3,d4,redner,red3,newman1,newman2,asb,alb2,alb3,havlin_r,Kertesz,Vespignani},
as it represents the largest publicly available computerized social
network.

Social networks have been much studied in social sciences
\cite{wasserman,kochen}. A general feature of these studies is that they are
restricted to rather small systems, and often view networks as
static graphs, whose nodes are individuals and links represent various quantifiable social interactions.

 In contrast,
recent approaches with methodology rooted in statistical physics focus on large networks,
searching for universalities both in
the topology of the web and in the dynamics governing it's evolution.
These combined theoretical and empirical results have opened
unsuspected directions for research and a wealth of applications in many fields
ranging from computer science to biology
\cite{watts2,asb,alb2,alb3,wasserman,alb4,lawrence,huberman,meta,sole1}.
Three important results seem to crystallize in this respect: First, most
networks have the the so called
{\em small world} property \cite{watts1,kochen}, which means that the average separation
between the nodes is rather small, i.e. one can find a short path along the links between most pairs of nodes.
	Second, real networks display a degree of
clustering higher than expected for random
networks\cite{watts1,watts2}.
	Finally, it has been found that the degree distribution contains
important information about the nature of the network, for many large networks
following a scale-free power-law distribution,
inspiring the study of {\em scale-free} networks
\cite{alb1,dorog1,d3,d4,redner,asb,alb2,alb3,sole1}.

In addition to uncovering generic properties of real networks, these studies
signal the emergence of a new set of modeling tools that considerably enhance
our ability to characterize and model complex interactive systems.
 To illustrate the power of this these advances
 we choose to investigate in detail the collaboration network of
scientists.

Recently Newman has taken an important step towards applying modern network ideas
to collaboration networks \cite{newman1,newman2}. He studied several large database
focusing on several fields of research over a five year period,
 establishing that collaboration networks
have all the general ingredients of small world networks: they have a
surprisingly short node-to-node distance and a large clustering coefficient\cite{newman1},
much larger than the one expected from a random Erd\H{o}s-R\'enyi
type network of similar size and average connectivity.
Furthermore, the degree distribution appears to follow a power law\cite{newman2}.

Our study takes a different, but complementary approach to collaboration networks than that followed by
Newman. We view collaboration networks as prototype of {\em evolving}
networks, where the accent is on dynamics and evolution. Indeed, the co-authorship network
constantly expands by the addition of new authors to the database, as
well as the addition of new internal links representing papers co-authored by authors that were
already part of the database. The topological properties of these
networks are determined by these dynamical and growth
processes. Consequently, in order to understand their topology,
we first need to understand the dynamical process that determines
their evolution. In this aspect Newman's study focuses on
the static properties of the collaboration graph, while our work investigates
 the dynamical properties of these networks. We show that such  dynamical
approach can explain many of the static topological features seen in the collaboration graph.

It is important to emphasize that the properties of the co-authorship network are not
unique. The WWW is also a complex evolving network, where nodes
and links are added (and removed) at a very high rate, the network topology being
profoundly
determined by these dynamical features\cite{alb1,alb4,lawrence,giles2}. The actor network of Hollywood is
very similar to the co-authorship network, because it grows through the addition of
new nodes (actors) and new links (movies linking existing actors)
\cite{watts1,watts2,alb3}.
Similarly, the nontrivial scaling properties of many cellular\cite{meta},
ecological\cite{sole1} or
business networks are all determined by dynamical processes that
contributed to the emergence of these networks. So
why single out the collaboration network as a case study? A number of factors
have contributed to this choice. First we needed a network for which the
dynamical evolution is explicitly available. That is, in addition to
a map of the network topology, it is important to know the time at
which the nodes and links have been added to the network,
crucial for revealing the network dynamics.
This requirement reduces the currently available databases to two systems:
the actor network, where we can follow the
dynamics by recording the year of the movie release, and the collaboration network for which the paper publication year
allows us to track the time evolution. Of these two, the
co-authorship
data is closer to a prototypical evolving network than the Hollywood actor
database for the following reasons: in the science collaboration network the
co-authorship decision is made entirely by the authors,
i.e. decision making is delegated to the level of individual nodes. In contrast, for actors
the decision often lies with the casting director, a level higher than the node.
While in the long run this difference is not particularly important, the
collaboration network is still closer in spirit to a prototypical
evolving network such as social systems or the WWW.

Our work stands on three pillars. First, we use direct measurements on the available
data to uncover the mechanism of network evolution. This implies determining
 the different parameters and uncovering the various competing
processes present in the system. Second, building on the mechanisms and
parameters revealed by the measurements we construct a model that allows us to
investigate the large scale topology the system, as well as its dynamical features.
The predictions offered by a continuum theory of the model allow us to
explain some of the results that were uncovered by ours,
as well Newman's measurements. The third and final step will involve computer
simulations of the model, serving several purposes: (i) It allows us to
investigate quantities that could not be extracted from the continuum theory;
(ii) Verifies the predictions of the continuum
theory; (iii) Allows us to understand the nature of the measurements we can perform on the
network, explaining some apparent discrepancies between the theoretical
and the experimental results.

\section{Databases: co-authorship in mathematics and neuro-science}

For each research field whose practitioners collaborate in
publications one can define a co-authorship
network which is a reflection of the professional links between the scientists.
In this network the nodes are the scientists
and two scientists are linked if they wrote a
paper together. In order to get information on the topology of a scientific
co-authorship web one needs a complete dataset of the published
papers, ideally from the birth of the discipline until today. However,
computer databases cover at most
the past several decades.
Thus any study of this kind needs to be limited to only a recent
segment of the database. This will
impose unexpected challenges, that need to be addressed, since such
limited data availability is a general feature of most networks.

The databases considered by us contain article titles and authors of all
relevant journals in the field of mathematics (M) and neuro-science (NS),
published in the period 1991-98. We have chosen these two fields
for several reasons. A first factor was the size of the database: biological
sciences or physics are orders of magnitude larger, too large to
address their properties with reasonable computing resources. Second, the selected
two fields offer sufficient diversity by displaying different publishing
patterns: in NS collaboration is intense, while mathematics, although
there is increasing tendency towards collaboration\cite{grossman}, is still a
basically single investigator field.

In mathematics our database contains 70,975 different authors and
70,901 papers for an interval spanning eight years. In NS the number of
different authors is 209,293 and the number of published papers is 210,750.
A complete statistics for the two considered database is summarized in
Fig.~1, where we plot the cumulative number of papers and
authors for the period 1991-98. We consider "new author"
an author who was not present in the database from 1991 up
to a given year.

%%%%%%%%%%%%%%%%%%%%%%%%%%%%%%%%%%%%%%%%%%%%%%%%%%%%%%%%%%%%%
\begin{figure}[h]
\begin{center}
\epsfig{figure=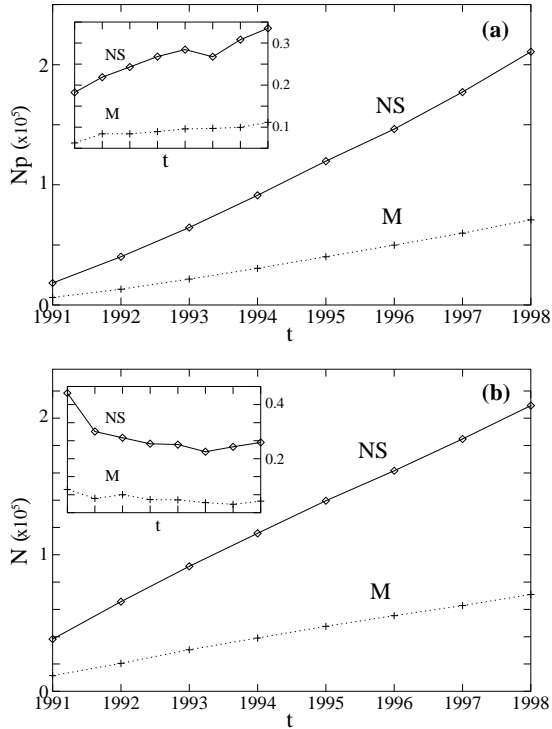,height=4.0in,width=3.0in}
\caption{ {\bf (a)} Cumulative number of papers for the M and NS databases
in the period 1991-98. The inset shows the number of papers published each year.
{\bf (b)} Cumulative number of authors (nodes) for the M and NS
databases in the period 1991-98.
The inset shows the number of new authors added each year.}
\label{fig:fig1}
\end{center}
\end{figure}
%%%%%%%%%%%%%%%%%%%%%%%%%%%%%%%%%%%%%%%%%%%%%%%%%%%%%%%%%%%%%
Before proceeding we need to clarify a few methodological
issues that affect the data analysis.
First, in the database the authors
are represented by their surname and initials of first and middle
name, thus there is a source of error in distinguishing some of them.
Two different authors with the same initials and surname will
appear to be the same node in the database.
This error is important mainly for scientists of Chinese and Japanese descent.
Second, seldom a given author uses one or two initials in different publications,
and in such cases he/she will appear as separate nodes.
Newman
\cite{newman1} showed that the error introduced by those problems
is of the order of a few percents. Our results are also affected by these
methodological limitations, but we do not expect that it will have a significant
impact on our results.

\section{Data analysis}

In this section we investigate the topology and dynamics of the
two databases, M and NS. Our goal is to extract the parameters
that are crucial to the understanding of the processes which
determine the network topology, offering input for the construction of an
appropriate model.

\subsection{Degree distribution follows a power-law}

\begin{figure}[h]
\begin{center}
\epsfig{figure=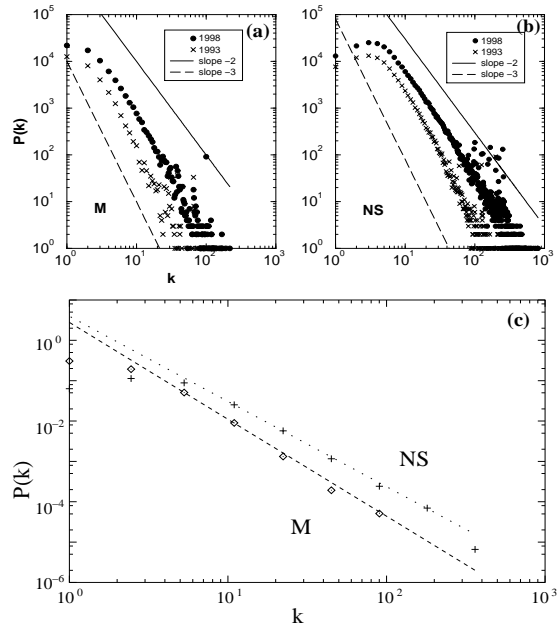,height=3.5in,width=3.0in}
\caption{ Degree distribution for the {\bf (a)} M and
{\bf (b)} NS database, showing the data based on the cumulative results up to yeas 1993 ($\times$)
and 1998 ($\bullet$).
{\bf (c)} Degree distribution shown with logarithmic binning
computed from the full dataset cumulative up to 1998. The lines correspond do the best fits,
and have the slope $2.1$ (NS, dotted) and $2.4$ (M, dashed).}
\label{fig:fig2}
\end{center}
\end{figure}

A quantity that has been much studied lately for various networks
is the degree distribution, $P(k)$, giving the probability that a
randomly selected node has $k$
links. Networks for which $P(k)$ has
a power-law tail, are known as {\em scale-free} networks\cite{alb1,alb2}.
On the other hand, classical network models, including the Erd\H{o}s-R\'enyi\cite{erdos,bollobas}
 and the
Watts and Strogatz\cite{watts2} models have an exponentially decaying $P(k)$ and are
collectively known
as {\em exponential} networks. The degree
distributions of both the M and NS data indicate that
collaboration networks are scale-free.
The power-law tail is evident from the raw, uniformly binned data (Fig.~\ref{fig:fig2}a,b), but
the scaling regime is better seen on the plot that uses
logarithmic binning,
reducing the noise in the tail (Fig.~2c). The
cumulative data with logarithmic binning indicates $\gamma_{M}=2.4$ and
$\gamma_{NS}=2.1$ for the two databases\cite{Shubert}.

%%%%%%%%%%%%%%%%%%%%%%%%%%%%%%%%%%%%%%%%%%%%%%%%%%%%%%%%%%%%%
%%%%%%%%%%%%%%%%%%%%%%%%%%%%%%%%%%%%%%%%%%%%%%%%%%%%%%%%%%%%%

We will see in the coming sections that the data indicates the existence of two
scaling regimes with two different scaling exponents.
The combination of these two regimes could easily give
the impression of an exponential cutoff in the $P(k)$ for large $k$.
Further analysis, offered in sections~\ref{continuum}-\ref{nonlin},
indicates that a consideration of two scaling regimes offers a more accurate
description.

\subsection{Average separation decreases in time}

\begin{figure}[h]
\begin{center}
\epsfig{figure=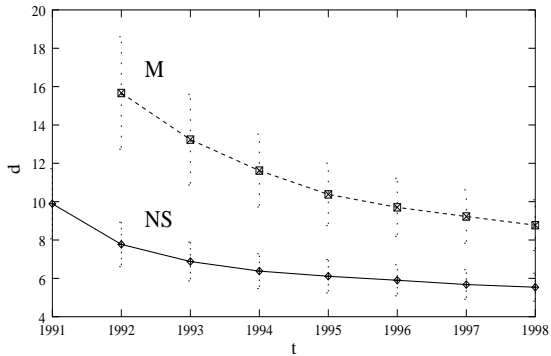,height=2.0in,width=3.0in}
\caption{Average separation in the M and NS databases.
The separation is computed on the cumulative data up to the indicated year.
The error bars indicate the standard deviation of the
distances between all pairs of nodes.
}
\label{fig:fig3}
\end{center}
\end{figure}

The ability of two nodes, $i$ and $j$, to communicate with each other depends
on the length of the shortest path, $l_{ij}$, between them. The average of
$l_{ij}$ over all pairs of nodes is denoted by $d=\,<l_{ij}>$,
and we call it the average separation of the network,
characterizing the networks interconnectedness. Large networks
can have surprisingly small separation, explaining the origin
of the {\em small-world} concept \cite{watts1,kochen}.
Determining the average separation in a large network is a rather time-consuming procedure.
Usually sampling
a fraction of all nodes and determining their distance from
all other points gives reasonable results. The results for the cumulative database are shown
in Fig.~\ref{fig:fig3}.

%%%%%%%%%%%%%%%%%%%%%%%%%%%%%%%%%%%%%%%%%%%%%%%%%%%%%%%%%%%%%
%%%%%%%%%%%%%%%%%%%%%%%%%%%%%%%%%%%%%%%%%%%%%%%%%%%%%%%%%%%%%

The figure indicates that $d$ decreases
with time, which is highly surprising because all network models so far predict
that the average separation should increase with system size\cite{alb4,bollobas}. The decreasing trend observed by us
could have two different origins. First, it is possible that indeed, the
separation does decrease as internal links, i.e. papers written by
authors that were previously part of the database, the
increase interconnectivity,thus decreasing  the diameter.
Second, the decreasing diameter could be a consequence of
the fact that we have no access to the full database, but only starting
from year 1991. As we demonstrate in sect.~\ref{szimulacio}, such incomplete dataset
could result in an apparently decreasing separation even if otherwise for the full
system the separation increases.

One can note the slow convergence of the diameter and the more
connected nature of the NS field expressed by a smaller separation.
The slow convergence
indicates that perhaps even longer time interval is needed to
reach the asymptotic limit, in which different relevant quantities take up a
stationary value.
The smaller separation for the NS field is expected, since
mathematicians tend to work in smaller groups and write papers with fewer
co-authors.

\subsection{Clustering coefficient decays with time}

\begin{figure}[h]
\begin{center}
\epsfig{figure=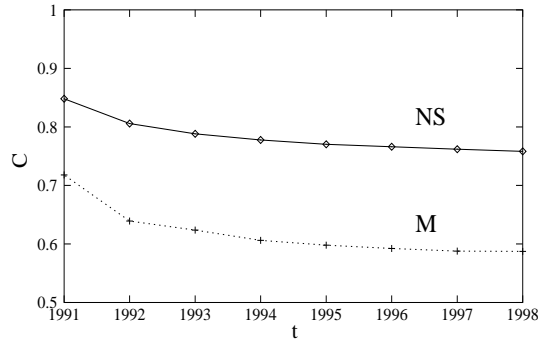,height=2.0in,width=3.0in}
\caption{Clustering coefficient of the M and NS database, determined for the
cumulative data up to the year indicated on the $t$ axis.}
\label{fig:fig4}
\end{center}
\end{figure}

An important phenomena characterizing the deviation of real networks from
the completely random ER model is clustering. The clustering
coefficient, a quantitative measure of this phenomena, $C$, can be defined
as follows\cite{watts2}:
pick a node, $i$ that has links to $k_i$ other nodes in the system.
If these $k_i$ nodes form a fully connected clique, there are $k_i(k_i-1)/2$ links
between them, but in reality we find much fever. Let us denote by $N_i$ the number of
links that connect the selected $k_i$ nodes to each other. The clustering
coefficient for node $i$ is then $C_i=2 N_i/k_i(k_i-1)$.
The clustering coefficient for the whole network is obtained by averaging
$C_i$ over all nodes in the system, i.e. $C=\,<C_i>_i$. In simple terms the
clustering coefficient of a node
in the co-authorship network tells us how much a node's collaborators
are willing to collaborate with
each other, and it represents the probability that two of it's
collaborators wrote a paper together.
The clustering coefficient for the
cumulative network as a function of time is shown in Fig.~\ref{fig:fig4}.
%%%%%%%%%%%%%%%%%%%%%%%%%%%%%%%%%%%%%%%%%%%%%%%%%%%%%%%%%%%%%
%%%%%%%%%%%%%%%%%%%%%%%%%%%%%%%%%%%%%%%%%%%%%%%%%%%%%%%%%%%%%

The results, in agreement with the separation measurements, suggest a stronger
interconnectedness for the NS
compared with M, and a slow convergence in time to an asymptotic value.

\subsection{Relative size of the largest cluster increases}

It is important to realize that the collaboration network is fragmented in many
clusters. There are several reasons for this. First, in every field there are
scientists that do not collaborate at all, that is they are the only authors of
all papers on which their name appears. This is more frequent in mathematics,
which despite
an increasing tendency toward collaboration \cite{grossman}, is
still more fragmented than physics or neural science. Second, and most
important,
the database contains papers published only after 1990. Thus there is a
possibility that two authors co-authored a paper before 1990, but in our database
they appear as disconnected.

\begin{figure}[h]
\begin{center}
\epsfig{figure=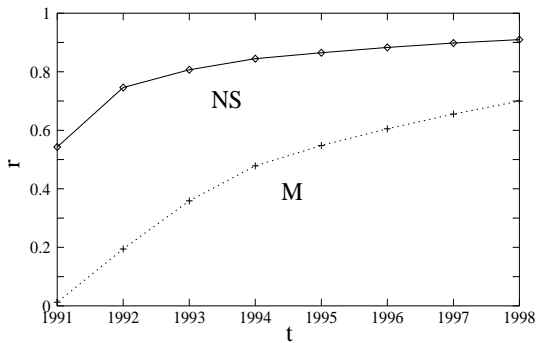,height=2.0in,width=3.0in}
\caption{Relative size of the largest cluster for the M and NS
database. Results are computed on the cumulative data up to the given year.}
\label{fig:fig5}
\end{center}
\end{figure}

If we look only at a single year, we see
many isolated clusters of authors. The cumulative dataset
containing several  years develops  a giant cluster, that contains a large
fraction of the authors. To investigate the emergence of this giant
connected component we measured the
relative size of the largest cluster, $r$, giving the ratio between the
number of nodes in the largest cluster and the total number of nodes in the
system. A cluster is defined as a subset of nodes interconnected by links.
Results from our cumulative co-authorship networks are presented in Fig.~\ref{fig:fig5}.
As expected, in M the fraction of clustered researchers is considerably
smaller than in NS.

%%%%%%%%%%%%%%%%%%%%%%%%%%%%%%%%%%%%%%%%%%%%%%%%%%%%%%%%%%%%%
%%%%%%%%%%%%%%%%%%%%%%%%%%%%%%%%%%%%%%%%%%%%%%%%%%%%%%%%%%%%%

The continuous increase in $r$ may  appear as the scenario commonly
described as percolation \cite{havlin} or the much studied emergence of the giant component
in random networks \cite{bollobas}. However, the process
leading to this giant cluster is
fundamentally different from these much studied phenomena. In most research fields, apart from a
very small fraction of authors that do not collaborate, all authors belong to a
single giant cluster from the very early stages of the field. That is,
the system is almost fully connected from the very first moment.
The only reason why the giant cluster
in our case grows so dramatically in the first several years is that we are
missing the information on the network topology before 1991. A good example is
the actor network, where the huge majority of the actors are part of the large
cluster at any stage of the network, starting from early 1900's until today.
However, if we would start recording collaborations only after 1990 for example,
the data would indicate, incorrectly, that many actors are disconnected.
The increasing $r$ indicates only the fact that we are reconstructing
the already existing giant cluster, and it is only a partial measure of it's emergence.

Finally, the fast convergence of the NS cluster size to an approximately
stationary value around 0.9 indicates that
after 1994 the network reached a
roughly stationary topology, i.e. the basic alliances are uncovered.
This does
not seems to be the case for M, where after ten years $r$ still
increases, perhaps due to smaller publication and collaboration rate in the
field.

\subsection{Average degree increases}

\begin{figure}[h]
\begin{center}
\epsfig{figure=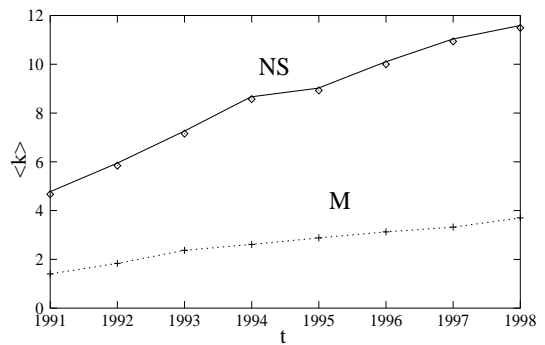,height=2.0in,width=3.0in}
\caption{Average number of links per node ($<k>$) for the M and NS
database. Results are computed on the cumulative data up to the given year.}
\label{fig:fig6}
\end{center}
\end{figure}

With time the number of nodes in our co-authorship network increases due to
arrival of new authors. The total number of links also increases through the
connections made by new authors with old ones and by new connections between
old authors. A quantity characterizing the network's interconnectedness
is the average degree $<k>$, giving the average number of links per author.
The time dependence of $<k>$ for
the cumulative network is shown in Fig.~\ref{fig:fig6}, indicating an approximately
linear increase of $<k>$ with time.
%%%%%%%%%%%%%%%%%%%%%%%%%%%%%%%%%%%%%%%%%%%%%%%%%%%%%%%%%%%%%
%%%%%%%%%%%%%%%%%%%%%%%%%%%%%%%%%%%%%%%%%%%%%%%%%%%%%%%%%%%%%
This is a rather important deviation from the majority of currently existing
evolving network models, that assume a constant  $<k>$ as the network expands.
As expected, the average degree  for M is much smaller than for NS.

\subsection{Node selection is governed by preferential attachment}\label{pref}

Classical network models assume complete randomness, i.e. the nodes are
connected to each other independent of the the number of links they already
had\cite{watts1,erdos,bollobas}.
The discovery of the power-law connectivity distribution required the
development of new modeling paradigms. A much used assumption is that in
scale-free networks nodes link with higher probability
to those nodes that already have a larger number of
links, a phenomena labeled as preferential attachment\cite{alb1,alb2}.
Implicitly or explicitly,
preferential attachment is part of all network models that aim to explain the emergence of the
inhomogeneous network structure and power law connectivity
distribution\cite{dorog1,d3,d4,redner}.
The availability of dynamic data on the
network development allows us to investigate its presence in the
co-authorship network. For this network preferential attachment
appears at two levels, that we discuss separately.

(i) {\em New nodes:} For a new author, that appears for the first time on a
publication, preferential attachment has a simple meaning: it is
more likely that the first paper will be co-authored with somebody that
already has a large number of co-authors (links) that with somebody
less connected. As a result "old" authors with more links will increase their
number of co-authors at a higher rate than those with fever links. To
investigate this process in quantitative terms we determined the probability
that an old author with connectivity
$k$ is selected by a new author for co-authorship. This probability
defines the $\Pi(k)$ distribution function. Calling "old authors"
those present up to the last year, and "new author" those who were added
during  the last year, we determine the change in the number of links, $\Delta k$,
for an old author that at the beginning of the last year had $k$ links.
Plotting $\Delta k$ as a function of $k$ gives the function $\Pi(k)$,
describing the nature of preferential attachment.
Since the measurements are limited to only a finite ($\Delta T=1$ year) interval,
we improve the statistics by plotting the integral of $\Pi(k)$:
\begin{equation}
\kappa(k)=\int_1^k \Pi(k')\, dk'.$$
\end{equation}

\begin{figure}[h]
\begin{center}
\epsfig{figure=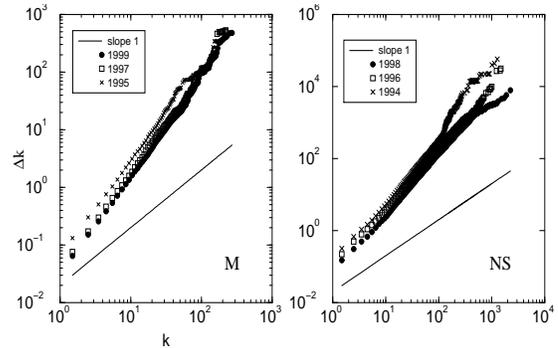,height=2.0in,width=3.0in}
\caption{Cumulated preferential attachment ($\kappa(k)$) of incoming new nodes for the
M and NS database. Results computed by considering
the new nodes coming in the specified year, and the network formed by nodes
already present up to this year. In the absence of preferential attachment
$\kappa (k)\sim k$, shown as continuous line on the figures.}
\label{fig:fig7}
\end{center}
\end{figure}

If preferential attachment is absent, $\Pi(k)$ should be independent of $k$,
as each node grows independently of it's degree, and $\kappa (k)$ is
expected to be linear. As Fig.~\ref{fig:fig7} shows, we find that
$\kappa (k)$ is nonlinear, increasing as $\kappa (k) \sim k^{\nu+1}$,
where the best fit gives $\nu \simeq 0.8$ for M and $\nu \simeq 0.75$ for NS.
This implies that $\Pi(k)\sim k^\nu$, where $\nu$ is different from
$1$ \cite{egyes}.
As simulations have shown, such nonlinear dependence generates deviations
from a power law $P(k)$\cite{egyes}. This was supported by analytical calculations\cite{redner},
that demonstrated that  the degree distribution follows a power law only for $\nu=1$.
The consequence of this nonlinearity will be discussed below.

%%%%%%%%%%%%%%%%%%%%%%%%%%%%%%%%%%%%%%%%%%%%%%%%%%%%%%%%%%%%%
%%%%%%%%%%%%%%%%%%%%%%%%%%%%%%%%%%%%%%%%%%%%%%%%%%%%%%%%%%%%%

\begin{figure}[h]
\begin{center}
\epsfig{figure=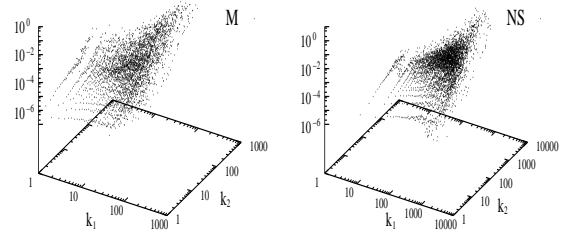,height=1.5in,width=3.2in}
\caption{Internal preferential attachment for the M and
NS database, 3D plots: $\Delta k$ as a function of $k_1$ and $k_2$.
Results computed on the cumulative data in the last considered year.}
\label{fig:fig8}
\end{center}
\end{figure}

(ii) {\em Internal links:} A large number of new links appear
between old nodes  as the network evolves, representing papers written by
authors that were part of the network, but did not collaborate before.
Such internal links are known to effect both the topology and
dynamics of the network \cite{dorog1}.
These internal links are also subject to
preferential attachment. We studied the probability $\Pi(k_1,k_2)$
that an old author
with $k_1$ links forms a new link with another old author with
$k_2$ links. The $\Pi(k_1,k_2)$ probability map
can be calculated by dividing $N(k_1, k_2)$, the  number of new links
between authors
with $k_1$ and  $k_2$ links, with the $D(k_1, k_2)$,
number of pairs of nodes with connectivities $k_1$ and $k_2$ present in the system:
\begin{equation}\label{1}
\Pi(k_1, k_2)=\frac{N(k_1,k_2)}{D(k_1,k_2)}.
\end{equation}
The three dimensional plot of $\Pi(k_1,k_2)$ is shown in Fig.\ref{fig:fig8}, the overall
behavior indicating preferential attachment: $\Pi(k_1,k_2)$
increases with as either $k_1$ or $k_2$'s increase.

%%%%%%%%%%%%%%%%%%%%%%%%%%%%%%%%%%%%%%%%%%%%%%%%%%%%%%%%%%%%
%%%%%%%%%%%%%%%%%%%%%%%%%%%%%%%%%%%%%%%%%%%%%%%%%%%%%%%%%%%%%

\begin{figure}[h]
\begin{center}
\epsfig{figure=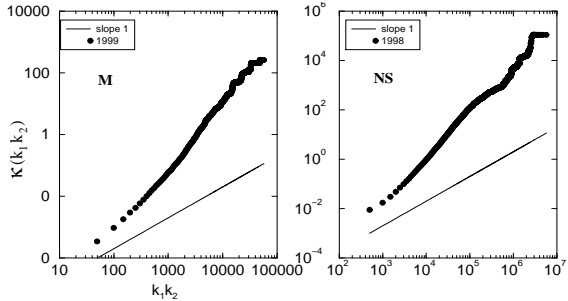,height=1.6in,width=3.0in}
\caption{Cumulated internal preferential attachment ($\kappa(k)$)
for the M  and
NS database, scaling as a function of the $k_1 k_2$ product.
Results computed on the cumulative data in the last considered year.
The straight lines have slope 1, expected for
$\kappa (k_1k_2)$ if there would be no preferential attachment.}
\label{fig:fig9}
\end{center}
\end{figure}

A natural hypothesis is to assume that $\Pi(k_1,k_2)$ factorizes into the
product $k_1k_2$. As Fig.~\ref{fig:fig9} shows, we indeed find that
\begin{equation}
\kappa(k_1k_2)=\int_1^{k_1k_2} \Pi(k_1'k_2')\, d(k_1'k_2')
\end{equation}
can be well approximated with a slope 2 as a function of $k_1k_2$,
indicating that for internal links the preferential attachment
is linear in the degree.

%%%%%%%%%%%%%%%%%%%%%%%%%%%%%%%%%%%%%%%%%%%%%%%%%%%%%%%%%%%%
%%%%%%%%%%%%%%%%%%%%%%%%%%%%%%%%%%%%%%%%%%%%%%%%%%%%%%%%%%%%%

\section{Modeling the web of science}\label{model}

In this section we use the obtained numerical results to construct a
simple model for the evolution of the co-authorship network.
It is important to emphasize that the
purpose of the model is to capture the main mechanisms that affect the evolution
and the scaling of the network, and not to incorporate every numerical detail of the
measured web. However,
the advantage of the proposed model is its flexibility: features,
neglected here, can be
incorporated into the current modeling framework.

We denote by $k_i(t)$ the
number of links node $i$ has at time $t$; by $T(t)$ and $N(t)$ the total number
of links and total number of nodes at time $t$, respectively. We
assume that all nodes present in the system are active, i.e. they can author
further papers. This is a reasonable assumption as the time-span over which data
is available to us is shorter than the professional lifetime of a scientist. In agreement with Fig.~\ref{fig:fig1},
we consider that new researchers
join the field at a constant rate, leading to
\begin{equation}\label{2}
N(t)=\beta t.
\end{equation}
The average number of links per node in
the system at time $t$ is thus given by:
\begin{equation}\label{3}
<k>\,=\frac{T(t)}{N(t)}.
\end{equation}
Fig.~\ref{fig:fig9} suggests, that the probability to create a new internal link
between two existing nodes is proportional with the product of their
connectivities. Consequently, denoting by $a$ the number of
newly created internal links per node in unit time, we write
the probability that between node $i$ and $j$ a new internal link
is created as
\begin{equation}\label{4}
\Pi_{ij}=\frac{k_i k_j}{\sum_{s,m}' k_s k_m}\, N(t)\, a,
\end{equation}
where the prime sign indicates that the
summation is done for $s\ne m$ values.

The measurements also indicated
(Fig.~\ref{fig:fig7}) that new nodes link to
the existing nodes with preferential attachment, $\Pi (k)$ follows $k^\nu$ with $\nu \simeq 0.75-0.8$.
Aiming to obtain an analytically solvable model, at this point we
neglect this nonlinearity and we approximate $\Pi (k)$ with a linear $k$
dependence. The effect of the nonlinearities will be discussed in sect.~\ref{nonlin}. Thus, if  node $i$ has $k_i$ links,
the probability that an incoming node will connect to it is
given by
\begin{equation}\label{5}
\Pi_i=b\,\frac{k_i}{\sum_{j} k_j},
\end{equation}
where $b$ is the average number of new links that an incoming node
creates.

We have thus formulated the dynamical rules that govern our evolving network
model, capturing the basic mechanism governing the evolution of the co-authorship network:
\begin{enumerate}
\item Nodes join the network at a constant rate.
\item Incoming nodes link to the already present nodes following
preferential attachment (\ref{5}).
\item Nodes already present in the network form new internal links following
 preferential attachment (\ref{4}).
\item We neglect the aging of nodes, and assume that all
nodes and links present in the system are active, able to initiate and
receive new links.
\end{enumerate}

 In the model we assume that the
 number of authors on a paper is constant. In reality $m$
is a stochastic variable, as the number of authors varies from paper to paper.
However, for the scale-free model the exponent $\gamma$
is known to be independent of $m$, thus making $m$ a stochastic variable is
not expected to change the scaling behavior.

\section{Continuum theory}\label{continuum}

Taking into account that new links join the system with a
constant rate, $\beta$,
the continuum equation for the
evolution of the number of links node $i$ has can be written as:
\begin{equation}\label{6}
\frac{dk_i}{dt}=\frac{b \beta k_i}{\sum_j k_j}+N(t)\, a \sum_{j}\,\!'
\frac{k_i k_j}{\sum_{s,m}' k_s k_m}.
\end{equation}
The first term on the right hand side
describes the contribution due to new nodes (\ref{5}) and the second term
gives the new links created with already existing nodes (\ref{4}). The total number of
links at time $t$ can be computed taking into account the
internal and external preferential attachment rules:
\begin{equation}\label{7}
\sum_i k_i=T(t)=\int_0^t 2\,[N(t')a+b \beta]\, dt'=t\beta (a t+2 b).
\end{equation}
Consequently the average number of links per node increases linearly in time,
\begin{equation}\label{8}
<k>\,=at+2b,
\end{equation}
in agreement with our measurements on the collaboration network (Fig.~\ref{fig:fig6}).
The master equation (\ref{6}) can be solved if we approximate the double sum
in the second term. Taking into account that we are interested in the
asymptotic limit where the total number of nodes is large relative to the
connectivity of the nodes,
we can write:
\begin{equation}\label{9}
\sum_{s,m}'k_s k_m=\sum_s k_s \sum_m k_m - \sum_m k_m^2 \approx
\left(\sum_i k_i\right)^2 .
\end{equation}
We have used here the fact that $T(t)^2$  depends on $N^2$, while
$\sum_i k_i^2$ depends only linearly on $N$ (we investigate the
$N\rightarrow \infty$ limit).
Using (\ref{9}) equation (\ref{6}) now becomes:
\begin{equation}\label{10}
\frac{dk_i}{dt}=\frac{b k_i}{t(at+2b)} + \frac{k_i a}{at+2b}.
\end{equation}
Introducing the notation $\alpha=a/b$, we obtain:
\begin{equation}\label{11}
\frac{dk_i}{dt}=\frac{k_i}{t}\,\frac{t\alpha+1}{t \alpha+2}.
\end{equation}
This differential equation is separable, the general
solution having the form
\begin{equation}\label{12}
k_i(t)=C_i \, \sqrt{t}\, \sqrt{2+\alpha t}.
\end{equation}
The $C_i$ integration constant can be determined from the initial conditions
for node $i$. Since node $i$ joins the system at time $t_i$,
we have $k_i(t_i)=b$, leading to
\begin{equation}\label{13}
k_i(t)=b\,\sqrt{\frac{t}{t_i}}\, \sqrt{\frac{2+\alpha t}{2+\alpha t_i}}.
\end{equation}
This implies that for large times ($t\rightarrow \infty$) the connectivity of the
node scales linearly with time, i.e. $k(t) \sim t$.

A quantity of major interest
is the degree distribution, $P(k)$.
The nodes join the system randomly at a constant rate,
which implies that the $t_i$ values are uniformly
distributed in time between 0 and t. The distribution function
for the $t_i$ in the $[0,t]$ interval is simply
$\rho(t)=1/t$. Using (\ref{13}), $P(k)$ can be obtained
after determining the $t_i(k_i)$ dependence from (\ref{13}), giving
\begin{eqnarray}\label{14}
P(k)&=&-\rho(t) \left.\frac{dt_i}{dk_i}\right|_{k}=\\
	&=& b^2(2/\alpha+t)\, \frac{1}{k^2}\,
\frac{1}{\sqrt{k^2+b^2t(2+\alpha t)}}.
\end{eqnarray}
An immediate consequence of this result is that the connectivity
distribution
depends both on the observation time $t$ and on the range of $k$
values we explore. In the asymptotic limit $t\rightarrow \infty$
we obtain
\begin{equation}\label{15}
P(k)\propto \frac{1}{k^2},
\end{equation}
predicting a scale-free behavior with exponent $\gamma=2$.
At short times, however, the exponent is different, the network
exhibiting a scale-free
behavior similar to the scale-free model \cite{alb1,alb2}:
\begin{equation}\label{16}
P(k)\propto \frac{1}{k^3}.
\end{equation}
Thus the model predicts that the degree distribution of the collaboration network displays
a crossover between two scaling regimes.
In general, scaling is controlled by the
time dependent crossover connectivity, given by
\begin{equation}\label{17}
k_c=\sqrt{b^2 t(2+\alpha t)}.
\end{equation}
For  $k\ll k_c$ the degree distribution scales
with an exponent $\gamma=2$, while for $k\gg k_c$ the degree distribution
scales with $\gamma=3$.
The crossover connectivity, $k_c$, increases linearly in time for $t \gg 2/\alpha$,
which implies that in the asymptotic limit ($t \rightarrow \infty$) only the
$\gamma=2$ exponent is observable.

Note that this result predicts that the
degree distribution has two scaling regimes, one with $\gamma=2$ for
small $k$, followed by a crossover to $\gamma=3$ for large $k$. This crossover
towards a larger exponent can be easily approximated with an exponential cutoff, which is why
we believe that in \cite{newman1} the power law with an exponential cutoff gave
a reasonable fit. However, as \cite{newman2} and our results show, for datasets with better
statistics the scaling regimes can be distinguished. Indeed,
the crossover is visible in Fig.~\ref{fig:fig2} as well, in  particular for
the degree distribution
of NS. The degree distribution taken in 1993 has a clear
$\gamma=3$ tail, as for the studied short time-frame (3 years) $k_c$
is expected to be low. This $\gamma=3$ tail all but disappears, however, in 1998, being
replaced with a $\gamma=2$ exponent, as predicted by (\ref{15}) for the limit $t\rightarrow \infty$.
The M database shows similar characteristics, albeit the crossover is masked by a higher
spread in the data point thanks to the weaker statistics.

\begin{figure}[h]
\begin{center}
\epsfig{figure=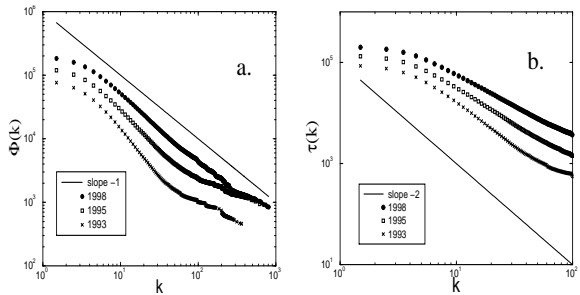,height=1.6in,width=3.0in}
\caption{Scaling of $\Phi(k)$ {\bf (a)} and
of $\tau(k)$ {\bf (b)} for the NS database, demonstrating the trends in the small and large
$k$ behavior of the degree distribution (see text).}
\label{fig:fig11}
\end{center}
\end{figure}

Plotting instead of $P(k)$ two differently cumulated values, the $\gamma=2$
and $\gamma=3$ scaling regimes are more evident.
Let us denote by $F(k)$ the primitive function of $P(k)$, defining:
\begin{equation}\label{18}
\Phi(k)=-F(1)-\int_1^k P(k')\, dk'.
\end{equation}
$\Phi(k)$ can be determined numerically by integrating $P(k)$ between $1$ and
$k$ and subtracting the constant at which the integral converges.
For small $k$ the function $\Phi(k)$ should scale as
\begin{equation}\label{19}
\Phi(k)\propto k^{-1},
\end{equation}
assuming that $P(k)$ scales as given by (\ref{15}).
As Fig.~\ref{fig:fig11}a shows, we indeed find that
for large $t$ (1998) the measured $\Phi (k)$ function
converges to a $k^{-1}$ behavior, which is less
apparent on the small $t$ curves (1993 and 1995).

To investigate the large $k$ behavior of $P(k)$ we measured the $\tau (k)$
function defined as:
\begin{equation}\label{20}
\tau(k)=\int_k^{\infty} P(k')\, dk',
\end{equation}
which captures the scaling of the tail. According to (\ref{16}) for large $k$ and small $t$
 one should observe
\begin{equation}
\tau(k)\propto k^{-2}.
\end{equation}
As Fig.~\ref{fig:fig11} shows, we indeed find that for NS for small $t$
(1993) the large $k$ scaling follows the prediction (\ref{20}), and,
as predicted, the scaling increasingly deviates from it as time increases.
%%%%%%%%%%%%%%%%%%%%%%%%%%%%%%%%%%%%%%%%%%%%%%%%%%%%%%%%%%%%
%%%%%%%%%%%%%%%%%%%%%%%%%%%%%%%%%%%%%%%%%%%%%%%%%%%%%%%%%%%%%

\section{Monte Carlo Simulations}\label{szimulacio}

While the continuum theory discussed in the previous section predicts the
connectivity distribution in agreement with the empirical data,
there are other quantities, such as the node
separation and clustering coefficient, that cannot be calculated using this
method at this
point. To investigate the behavior of these measures of the network
topology next we study the model proposed in Sect.~\ref{model} using Monte Carlo
simulations.

Due to memory and
computing time limitations we investigated relatively small networks,
with total number of nodes $N<4000$. While these networks are
considerably smaller than the real networks, their scaling and
topological features should be representative.
In order to form a reasonable number of internal links, we increased the
parameter $a$ in Eq.~(\ref{4}). For comparison purposes we note that in the real system we have
$a_{\text{M}}=0.31/{\text{year}}
\simeq 10^{-4}/{\text{simulation step}}$ and $a_{\text{NS}}=0.98/{\text{year}}
\simeq 3.684\cdot10^{-5}/\text{simulation step}$, numbers that can be derived from the data shown in
Fig.~\ref{fig:fig6} and Fig.~\ref{fig:fig1}b.

The advantage of the modeling efforts, including the Monte Carlo simulations,
is that they reproduce the network dynamics from the very first node.
In contrast, the database we studied records nodes and links only after 1991, when
much of the networks structure was already in place.
By collecting data over several
years we gradually discovered the underlying structure.
We expect that after a quite
long measurement time the structure revealed by the collected data
will be statistically indistinguishable from the full collaboration network.
 However, the dynamics we measure during this process for the relevant
quantities (diameter, average connectivity, clustering coefficient) might
differ from those characterizing the full network,
since all of them are computed on the {\em incomplete} network
(revealed by the available data). However, Monte Carlo simulations allow us to
investigate the effect of the data incompleteness on the relevant network
measures.

We investigated the time dependence of
the average connectivity, the diameter and
the clustering coefficient, using the parameters
$N_{max}=1000$, $a=0.001$,
$\beta=1$ and $b=2$. In order to increase the statistics, the results were
averaged over 10 independent configurations.

\begin{figure}[h]
\begin{center}
\epsfig{figure=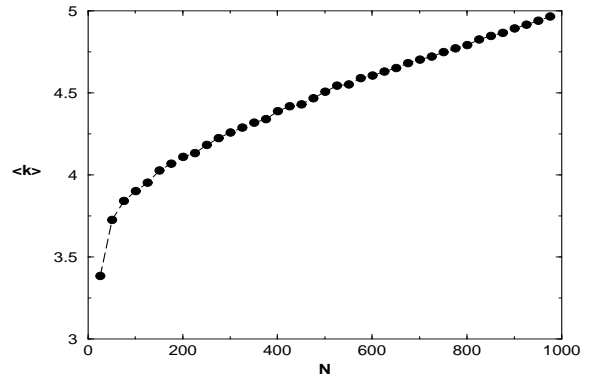,height=2in,width=3.0in}
\caption{Computer simulated dynamics of the average connectivity
in the proposed model. ($N_{max}=1000$, $a=0.001$, $\beta=1$ and $b=2$)}
\label{fig:fig13}
\end{center}
\end{figure}

\emph{Average degree:}  As Fig.~\ref{fig:fig13} indicates, asymptotically the average connectivity
increases linearly, in agreement with both our measurements (see Fig.~\ref{fig:fig6}) and the
continuum theory (see Eq.~(\ref{3})).
%%%%%%%%%%%%%%%%%%%%%%%%%%%%%%%%%%%%%%%%%%%%%%%%%%%%%%%%%%%%
%%%%%%%%%%%%%%%%%%%%%%%%%%%%%%%%%%%%%%%%%%%%%%%%%%%%%%%%%%%%%

\emph{Average separation:} The empirical results indicated (see Fig.~\ref{fig:fig3})
that the average separation decreases with time for both databases. In contrast,
our simulations show a monotonically increasing $d$, in apparent disagreement with the real system.
%%%%%%%%%%%%%%%%%%%%%%%%%%%%%%%%%%%%%%%%%%%%%%%%%%%%%%%%%%%%
\begin{figure}[h]
\begin{center}
\epsfig{figure=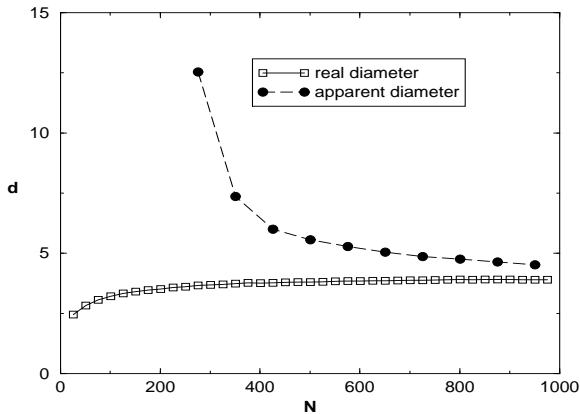,height=2.2in,width=3.0in}
\caption{Computer simulated dynamics for the real and apparently
measured diameter value.
($N_{max}=1000$, $a=0.001$, $\beta=1$, $b=2$ and $N_s=200$)}
\label{fig:fig14}
\end{center}
\end{figure}
%%%%%%%%%%%%%%%%%%%%%%%%%%%%%%%%%%%%%%%%%%%%%%%%%%%%%%%%%%%%%

Note that an increasing diameter agrees with measurements done on
other models, including
scale-free and exponential networks, that all predict an approximately logarithmic increase
with the number of nodes, $d\propto \ln(N)$\cite{bollobas,diameter}.
This contradiction between the models and our empirical data is rooted in the
incomplete data we have for the first years of our measurements.
To show this we perform the following simulation. We construct a network of $N=1000$
nodes. However, we will record the apparent diameter of the network made of nodes that have been
added only after a predefined time,
mimicking the fact that the data available for us gives $d$ only for publications after 1991.
We find that the separation of this incomplete  network has a decreasing tendency,
slowly converging to the real value (Fig.~\ref{fig:fig14}),
in agreement with the decrease observed in the empirical measurements (Fig.~\ref{fig:fig3}).
This result underlies the importance of simulations in understanding the dynamics of
complex networks, and resolves the conflict between the simulation and the empirical data.
It also indicates that most likely the diameter of the M and NS database does increase in
time, but such increase can be observed only if much longer
time intervals will be available for study.

%%%%%%%%%%%%%%%%%%%%%%%%%%%%%%%%%%%%%%%%%%%%%%%%%%%%%%%%%%%%
\begin{figure}[h]
\begin{center}
\epsfig{figure=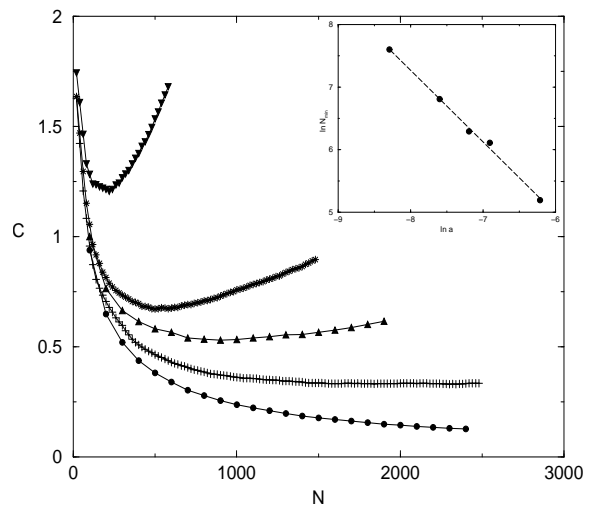,height=2.7in,width=3.0in}
\caption{Clustering coefficient for different
values of the $a$ parameter as a function of the system size $N$.
($N_{max} = 1000$, $\beta=1$ and $b=2$, values of $a$ are 0 ($\bullet$),
0.00025 ($+$), 0.0005 ($\bigtriangleup$), 0.00075 ($\ast$), 0.002 ($\bigtriangledown$).)
The inset shows the scaling of the $N_{min}$ value as a function of the
$a$ parameter. ($N_{max}=1000$, $\beta=1$ and $b=2$, the line shows a fit
$\ln N_{min}=-1.887-1.144\cdot \ln a$.)
}
\label{fig:fig15}
\end{center}
\end{figure}
%%%%%%%%%%%%%%%%%%%%%%%%%%%%%%%%%%%%%%%%%%%%%%%%%%%%%%%%%%%%%

\emph{Clustering coefficient.}
The clustering coefficient predicted by our simulations is shown on Fig.~\ref{fig:fig15}.
As the figure indicates, C depends strongly
on the value of the parameter $a$. For $a=0$ we have essentially the scale-free
model \cite{alb1} and the clustering coefficient has a monotonically decreasing tendency.
For $a>0$ however, the clustering coefficient decreases at
the beginning and
after reaching a minimum at $N_{min}$ changes its course, asymptotically
increasing with time.
Thus, for all $a>0$, we expect that in the
asymptotic limit the clustering coefficient should increase, in
agreement with our measurements on the collaboration network (see Fig.~\ref{fig:fig4}). The $N_{min}$
position where the clustering coefficient has a minimum
scales as power of the $a$ parameter, as shown as the inset in Fig.~\ref{fig:fig15}.
%%%%%%%%%%%%%%%%%%%%%%%%%%%%%%%%%%%%%%%%%%%%%%%%%%%%%%%%%%%%
%\begin{figure}[h]
%\begin{center}
%\epsfig{figure=fig18.eps,height=2in,width=3.0in}
%\caption{Scaling of the $N_{min}$ value as a function of the
%$a$ parameter. ($N_{max}=1000$, $\beta=1$ and $b=2$)}
%\label{fig:fig16}
%\end{center}
%\end{figure}
%%%%%%%%%%%%%%%%%%%%%%%%%%%%%%%%%%%%%%%%%%%%%%%%%%%%%%%%%%%%%

We conclude thus that the decreasing $C$
observed for our database, shown in Fig.~\ref{fig:fig4}, does not represent the asymptotic
behavior. The observed behavior also indicates that one should view the values for
$C$ reported in the literature, and measured for finite time-frames (maximum 5
years) with caution, as they might not represent asymptotic values.

\begin{figure}[h]
\begin{center}
\epsfig{figure=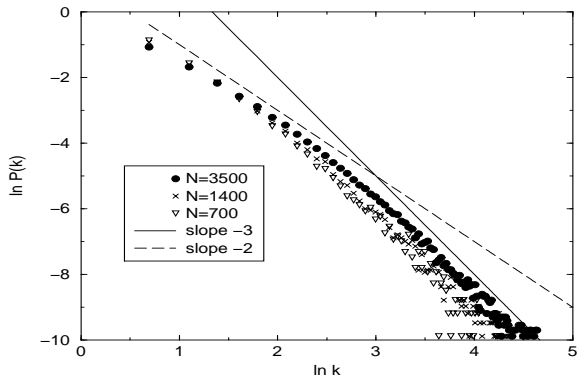,height=2in,width=3.0in}
\caption{Connectivity distributions as predicted by numerical simulation
for different stages of evolution of the network
($a=0.001$, $\beta=1$ and $b=2$).}
\label{fig:fig17}
\end{center}
\end{figure}

\emph{Degree distribution:}
The simulations provide $P(k)$ as well, allowing us to check the validity
of the predictions of the continuum theory.
 Although the considered system sizes are rather small
($N_{max}=3500$)
compared to the $N\rightarrow \infty$ approximation used in the analytical
calculation and the $N_M=70,975$, $N_{NS}=209,750$ for the empirical data,
the behavior of $P(k)$, shown in Fig.~\ref{fig:fig17}
agrees with our continuum model and
measurements. For small $k$ we observe the $\gamma=2$ scaling, while for
 large $k$ $P(k)$ converges to the predicted $\gamma=3$ exponent.

%%%%%%%%%%%%%%%%%%%%%%%%%%%%%%%%%%%%%%%%%%%%%%%%%%%%%%%%%%%%
%%%%%%%%%%%%%%%%%%%%%%%%%%%%%%%%%%%%%%%%%%%%%%%%%%%%%%%%%%%%%

\section{Nonlinear effects}\label{nonlin}

	An issue that remained unresolved up to this pont concerns the effect
of the nonlinear preferential attachment. We have seen in sect.~\ref{pref}
that for the incoming links we have
\begin{equation}
\Pi_i=b\, \frac{k_i^{\nu}}{\sum_{j} k_j^{\nu}},
\end{equation}
with $\nu \approx 0.8$. On the other hand, for such preferential attachment
Krapivsky \emph{et al} have shown that the degree distribution
follows a stretched exponential, i.e. the power law is absent\cite{redner}. This
would indicate that $P(k)$ for the co-authorship network should follow a
stretched exponential, which disagrees with our and Newman's findings (we have
explicitly checked that a stretched exponential is not a good fit for our data).
What could then override the known effect of the $\nu<1$ nonlinear behavior?
	Next we propose a possible explanation: the linearity of the internal preferential attachment
can restore the power law nature of $P(k)$.

For non integer $\nu$ values the differential equation
(\ref{6}) governing the evolution of the connectivity is not analytically
solvable. However, in the extreme case $\nu=0$ (no preferential
attachment for new nodes) the equation is again analytically tractable.
Equation (\ref{6}) in this case has the form
\begin{equation}
\frac{dk_i}{dt}=\frac{b \beta}{N(t)} + N(t)\, a \sum_j \frac{k_i k_j}{\sum_{s,m}
k_s k_m}.
\end{equation}
Using  $N(t)=\beta t$ and $<\!k\!>\,\,=at+2b$, which are valid in this
case as well, following the steps described
in sect.~\ref{continuum}, we obtain the
differential equation:
\begin{equation}
\frac{dk_i}{dt}=\frac{b}{t}+\frac{a\,k_i}{at+2b}.
\end{equation}
The general solution of this equation has the form:
\begin{equation}
k_i(t)=(2b+at)\,C_i+\frac{1}{2} (2b+at) \log\left(\frac{t}{2b+at}\right) ,
\end{equation}
where $C_i$ is an integration constant which can be determined
using the $k_i(t_i)=b$ initial condition. We thus obtain:
\begin{equation}
k_i(t)=b\,\frac{2b+at}{2b+at_i} + \frac{1}{2} (2b+at) \log\left[\frac{t(2b+at_i)}{t_i(2b+at)}\right]
\end{equation}
The degree distribution cannot be determined analytically, since the
$t_i(k_i)$ function is not analytical. However, taking the $\{ t, t_i\}
\rightarrow \infty$ limit, i.e. focusing on the network's long time evolution
we obtain
\begin{equation}
k_i(t)\approx b\,\frac{t}{t_i},
\end{equation}
which again predicts a power-law degree distribution:
\begin{equation}
P(k) \propto \frac{1}{k^2}.
\end{equation}

Consequently, we obtain that in
the asymptotic limit for $\nu=0$ the scale-free degree
distribution has the same tail as we obtained for $\nu=1$.
This result suggests that the linearity in the
internal preferential attachment determines the asymptotic form of the degree
distribution. The real exponent $\nu=0.8$ is between the two asymptotically
solvable cases $\nu=0$ and $\nu=1$, but, based on the limiting behavior of the two
extremes we
expect that independently of the  value of
$0 \leq \nu \leq 1$, in the asymptotic limit the degree distribution should
converge to a power-law with $\gamma=2$.
On the other hand, we expect that the nonlinear $\nu \neq 1$ behavior would have a
considerable effect on the non-asymptotic behavior, which is not accessible analytically at this point.

\begin{figure}[h]
\begin{center}
\epsfig{figure=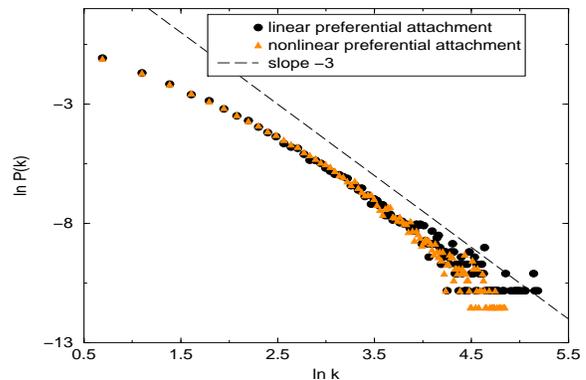,height=2in,width=3.0in}
\caption{Connectivity distribution generated by the numerical simulations
for linear ($\nu=1$) and nonlinear ($\nu=0.75$) preferential attachment
($N_{max}=3500$, $a=0.0005$, $\beta=1$ and $b=2$).}
\label{fig:szimul}
\end{center}
\end{figure}

To test further the potential effect of the nonlinearities, we have simulated the
model discussed in sect.~\ref{szimulacio} with $\nu=0.75$,
otherwise all parameters being unchanged. We show on Fig.~\ref{fig:szimul} the degree distribution
for the linear ($\nu=1$) and the nonlinear ($\nu=0.75$) case.
%%%%%%%%%%%%%%%%%%%%%%%%%%%%%%%%%%%%%%%%%%%%%%%%%%%%%%%%%%%%
%%%%%%%%%%%%%%%%%%%%%%%%%%%%%%%%%%%%%%%%%%%%%%%%%%%%%%%%%%%%%
As one can see, the $\nu=1$ and $\nu=0.75$ case can be hardly
distinguished. This could have two origins. First, the simulations are limited to
$t=3500$ simulation steps, due to the discussed running time limitations.
Thus we are hardly in the asymptotic
regime. On the other hand, the agreement indicates that the nonlinear effect has hardly distinguishable
effect on $P(k)$, again the internal attachment dominating the system behavior.

	In summary, the domination of the internal attachment affects are expected to be even more
dominant for the real network.
	Indeed, in the collaboration network the fraction of the links created as
internal links is much higher than those created by the incoming nodes, as
 an author qualifies
for a new incoming link only on his first paper. Most scientists contribute for a considerable time to the
same filed, publishing numerous subsequent papers, and these later links will all appear as internal links.
Thus typically the number of internal links is much higher than the number of new links, the
network's topology is much more driven by the internal
links then by the external ones. This is one possible reason why the effect of the nonlinear behavior,
while clearly present, cannot be detected in the functional form of $P(k)$.

\section{Discussion}

In the last two years we witnessed considerable advances in addressing the topology and the dynamics of
complex networks. Along this road a number of quantities have been measured and calculated,
aiming to characterize the network topology. However most of these studies are fragmented, focusing
on one or a few characteristics of the network at a time.
	Here we presented a detailed study of a network of high
interest to the scientific community, the collaboration network of scientists, which also
represents a prototype example of a complex evolving network.
	This study allows us to investigate
to which degree can we use various known measures to characterize a given
network. A first and important result of our investigation
is that we need to be careful at distinguishing between
the asymptotic and the intermediate behavior. In particular, most
quantities used to characterize the network are time
dependent. For example, the diameter, the clustering coefficient,
as well as the average degree of the nodes are
often used as basic time independent
network characteristics. Our empirical results show that many of these key quantities are time dependent,
without a tendency to saturate within the available time-frame.
Thus their value at a given moment tells us little about the network.
They can be used, however, at any moment, to show that the network has small world properties, i.e. it
has a small average separation, and a clustering coefficient that is larger than one expected
for a random network.

	A quantity that is often believed to offer a stationary measure of the network is the
degree distribution. Our empirical data, together with the analytic solution of the model
shows that this is true only asymptotically for the co-authorship network: we uncover a crossover behavior
between two different scaling regimes. We tend to believe that the model's predictions are
not limited to the collaboration network: as on the WWW and for the actor collaboration network similar
basic processes take place, chances are that similar crossovers could appear there as well.

	A third important conclusion of the study regards the understanding that the measurements done on
incomplete databases could offer trends that are opposite compared to that seen in the full system.
An example is the node separation: we find that the empirically observed
decreasing tendency is an artifact of the incomplete data.
However, our simulations show that one can, with careful modeling, uncover such inconsistencies.
But this also offers an important warning: for any network, before attempting to model it, we need to fully
understand the limitations of the data collection process, and test their effect on the
quantities of interest for us.

	The model presented here represents only the starting point toward a complete modeling of
collaborations in science. As we discussed thorough the paper, we have made several
important approximations, sacrificing certain known network features for an analytical solution.
For example, we neglected in our modeling effort the potential effect of aging \cite{d3,asb},
reflecting the fact that scientists retire and stop publishing papers.
	In the long run such aging effects will, undoubtedly, introduce exponential cutoffs in $P(k)$,
as there are inherent limits on how many papers a researcher can write. Those
effects, however, are not visible in our datasets. There are several potential
reasons for this.
	Probably the most important is that even the eight years available to us for study is much
shorter than the professional life of a scientist. Such aging induced cutoffs are expected to be visible only
when time-frames of length of several time the scientist's professional life are studied. Data
availability so far does not permit such studies.
	
	A second simplification is that we assumed that each paper has exactly $m$ authors.
	That is far from being so, as the numbers of co-authors varies greatly between papers.
	However, it is hard to imagine that the inclusion of a stochastic component in $m$ would
fundamentally affect our results.
	It is clear that such stochastic component will not affect $P(k)$, and we feel
that the effect on $d$ or $C$ is also negligible, but we lack at this point results to support
this latter claim.

	A surprising result of our study is the power law character of $P(k)$, despite the fact that
$\Pi(k)$ is nonlinear.
	We have shown that the existence of a linear internal attachment rule is able
to restore the power law $P(k)$.
	Considering the fact that the largest fraction of links appear as internal links, compared
with links created by new authors, it is fair to expect that the scaling determined by this
internal linking process will dominate.
	The fact that for the two limits of the internal linking exponents, $\nu=0$ and $\nu=1$,
we obtained power law $P(k)$ despite the nonlinear external $\Pi(k)$, suggests that
such power law might appear for nonlinear, $\nu\neq 0,1$ internal $\Pi(k)$ as well.
	Solving this problem is a formidable challenge, but it is perhaps worth the effort.

Finally, a more detailed modeling of the co-authorship network would involve the
construction of bipartite graphs \cite{bipart}, in which we directly simulate the publishing of papers
by several co-authors, which are all connected to each other.
In such a model the basic unit is a paper, that involves several "old" and "new" authors.
In such a framework one can simultaneously study the evolution of the co-authorship network
(in which nodes are scientists linked by joint publications) and the publication network
(in which nodes are papers linked by joint authors).
	One can imagine that coupled continuum equations could be formulated for such bipartite
network as well, which would eventually predict the network's dynamics and topology.
	Undoubtedly including such detail in the modeling  effort would increase the fidelity of the model.
While challenging, following such path is beyond our goals here.

	In summary, the modeling efforts presented here are only the starting point for a systematic
investigation of the evolution of social networks. It is important to note that such modeling
is open ended: more details can be incorporated, that could undoubtedly improve the agreement
between the empirical data and theory.
	And such improvements might not be in vain: they could point towards a better understanding
of the evolution of not only the co-authorship graph, but complex networks in general.

\section{Acknowledgments}

This work has been done during a collaboration at the Collegium Budapest
Institute of Advanced Study, Hungary. We gratefully acknowledge the inspiring
and professionally motivating atmosphere, and the fellowships
offered by Collegium Budapest in supporting this study. We thank M. Newman
and I. Der\'enyi for discussions and comments on this topic.

%:::::::::::::::::::::::::::::::::::::::::::::::::::::::::::::::::::::::
% References:
%:::::::::::::::::::::::::::::::::::::::::::::::::::::::::::::::::::::::

%:::::::::::::::::::::::::::::::::::::::::::::::::::::::::::::::::::::::
{\bf Figure Captions:}
%:::::::::::::::::::::::::::::::::::::::::::::::::::::::::::::::::::::::

{\bf Fig.~1}  {\bf (a)} Cumulative number of papers for the M and NS databases
in the period 1991-98. The inset shows the number of papers published each year.
{\bf (b)} Cumulative number of authors (nodes) for the M and NS
databases in the period 1991-98.
The inset shows the number of new authors added each year.

{\bf Fig.~2}  Degree distribution for the {\bf (a)} M and
{\bf (b)} NS database, showing the data based on the cumulative results up to yeas 1993 ($\times$)
and 1998 ($\bullet$).
{\bf (c)} Degree distribution shown with logarithmic binning
computed from the full dataset cumulative up to 1998. The lines correspond do the best fits,
and have the slope $2.1$ (NS, dotted) and $2.4$ (M, dashed).

{\bf Fig.~3} Average separation in the M and NS databases.
The separation is computed on the cumulative data up to the indicated year.

{\bf Fig.~4} Clustering coefficient of the M and NS database, determined for the
cumulative data up to the year indicated on the $t$ axis.

{\bf Fig.~5} Relative size of the largest cluster for the M and NS
database. Results are computed on the cumulative data up to the given year.

{\bf Fig.~6} Average number of links per node ($<k>$) for the M and NS
database. Results are computed on the cumulative data up to the given year.

{\bf Fig~7}  Cumulated preferential attachment ($\kappa(k)$) of incoming new nodes for the
M and NS database. Results computed by considering
the new nodes coming in the specified year, and the network formed by nodes
already present up to this year. In the absence of preferential attachment
$\kappa (k)\sim k$, shown as continuous line on the figures.

{\bf Fig.~8} Internal preferential attachment for the M and
NS database, 3D plots: $\Delta k$ as a function of $k_1$ and $k_2$.
Results computed on the cumulative data in the last considered year.

{\bf Fig.~9} Cumulated internal preferential attachment ($\kappa(k)$)
for the M  and
NS database, scaling as a function of the $k_1 k_2$ product.
Results computed on the cumulative data in the last considered year.
The straight lines have slope 1, expected for
$\kappa (k_1k_2)$ if there would be no preferential attachment.

{\bf Fig.~10} Scaling of $\Phi(k)$ {\bf (a)} and
of $\tau(k)$ {\bf (b)} for the NS database, demonstrating the trends in the small and large
$k$ behavior of the degree distribution (see text).

{\bf Fig.~11} Computer simulated dynamics of the average connectivity
in the proposed model. ($N_{max}=1000$, $a=0.001$, $\beta=1$ and $b=2$)

{\bf Fig.~12} Computer simulated dynamics for the real and apparently
measured diameter value.
($N_{max}=1000$, $a=0.001$, $\beta=1$, $b=2$ and $N_s=200$)

{\bf Fig.~13} Clustering coefficient for different
values of the $a$ parameter as a function of the system size $N$.
($N_{max} = 1000$, $\beta=1$ and $b=2$, values of $a$ are 0 ($\bullet$),
0.00025 ($+$), 0.0005 ($\bigtriangleup$), 0.00075 ($\ast$), 0.002 ($\bigtriangledown$).)
The inset shows the scaling of the $N_{min}$ value as a function of the
$a$ parameter. ($N_{max}=1000$, $\beta=1$ and $b=2$, the line shows a fit
$\ln N_{min}=-1.887-1.144\cdot \ln a$.

{\bf Fig.~14} Connectivity distributions as predicted by numerical simulation
for different stages of evolution of the network
($a=0.001$, $\beta=1$ and $b=2$).

{\bf Fig.~15} Connectivity distribution generated by the numerical simulations
for linear ($\nu=1$) and nonlinear ($\nu=0.75$) preferential attachment
($N_{max}=3500$, $a=0.0005$, $\beta=1$ and $b=2$).

\end{document}